\documentclass[pre,citeautoscript,superscriptaddress,twocolumn,nofootinbib,floatfix]{revtex4-1}
\usepackage{xcolor}
\usepackage{graphicx}
\usepackage{bbm}
\usepackage[utf8]{inputenc}
\usepackage[T1]{fontenc}
\usepackage{amsmath,amssymb,amsthm}

\newcommand{\V}{\mathcal{V}}
\newcommand{\Q}{\mathcal{Q}}
\newcommand{\Sur}{\mathcal{S}}
\newcommand{\Sig}{\mathcal{Z}}
\newcommand{\midd}{\parallel}
\DeclareMathOperator*{\argmax}{arg\,max}

\usepackage{algpseudocode}

\definecolor{link-color}{cmyk}{0.8 ,  0.3 ,  0. , 0}
\usepackage[colorlinks=true,%
            linkcolor   = {link-color},%
            citecolor   = {link-color},%
            urlcolor    = {link-color}]{hyperref}

\usepackage[caption=false]{subfig}

\graphicspath{ {../figures/} {figures/} }
\begin{document}

  \author{V.A. Traag}
  \email{traag@kitlv.nl}
  \affiliation{Royal Netherlands Institute of Southeast Asian and Caribbean Studies, Leiden}
  \affiliation{e-Humanities Group, Royal Netherlands Academy of Arts and
  Sciences, Amsterdam}
  \author{R. Aldecoa}
  \affiliation{Department of Physics, Northeastern University, Boston}
  \author{J-C. Delvenne}
  \affiliation{ICTEAM, Universit\'e catholique de Louvain, Louvain-la-Neuve}

  \title{Detecting communities using asymptotical surprise}

  \begin{abstract}

  Nodes in real-world networks are repeatedly observed to form dense clusters, often referred to as communities. 
  Methods to detect these groups of nodes usually maximize an objective function, which implicitly contains the definition of a community. 
  We here analyze a recently proposed measure called surprise, which assesses the quality of the partition of a network into communities. 
  In its current form, the formulation of surprise is rather difficult to analyze.
  We here therefore develop an accurate asymptotic approximation. 
  This allows for the development of an efficient algorithm for optimizing surprise.
  Incidentally, this leads to a straightforward extension of surprise to weighted graphs. 
  Additionally, the approximation makes it possible to analyze surprise more closely and compare it to other methods, especially modularity.
  We show that surprise is (nearly) unaffected by the well known resolution limit, a particular problem for modularity.
  However, surprise may tend to overestimate the number of communities, whereas they may be underestimated by modularity.
  In short, surprise works well in the limit of many small communities, whereas modularity works better in the limit of few large communities.
  In this sense, surprise is more discriminative than modularity, and may find communities where modularity fails to discern any structure.
  \end{abstract}

\maketitle

\section{Introduction}

Networks are often used as a model to describe interactions among components of
a system~\cite{albert_statistical_2002,dorogovtsev_lectures_2010}. In its
simplest form, a network is composed of a set of vertices (also called nodes)
and a set of edges connecting them.  Many real-world systems can be reduced to
this scheme, such as social networks establishing relations among individuals,
proteins interacting within the cell or roads connecting different
cities~\cite{newman_networks:_2010}. What caught the interest of the scientific
community was that most of these real networks share high-order structural
patterns and dynamics, such as a wide heterogeneity in the number of neighbors
of a node, the presence of many triangles or a very low network
diameter~\cite{barabasi_emergence_1999,watts_collective_1998}.  Another feature
observed in real networks is the presence of densely connected groups of nodes,
known as communities~\cite{fortunato_community_2010}. 
Nodes in the same group usually share similar characteristics or functions and, therefore, methods to detect communities in networks are of much interest across different fields~\cite{girvan_community_2002,gleiser_community_2003,palla_uncovering_2005,guimera_worldwide_2005,olesen_modularity_2007,lupu_trading_2013}

Researchers have proposed numerous strategies to detect the community structure of a network~\cite{danon_comparing_2005,lancichinetti_finding_2011,fortunato_community_2010,aldecoa_surprise_2013}.
Ultimately, most methods optimize a given objective function to find a partition into communities.
This function contains, either explicitly or implicitly, its own definition of a community. 
Modularity~\cite{newman_finding_2004} has been, since its inception,
the most extensively used measure for community detection. It belongs to a wider
class of functions in which communities are defined by Potts model spin states
and the quality of the partition is given by the energy of the
system~\cite{reichardt_statistical_2006,ronhovde_local_2010}.  Although this
approach based on statistical mechanics may be appealing, empirical evidence
shows that in many cases these methods are unable to capture the expected
communities of the
network~\cite{lancichinetti_community_2009,good_performance_2010,aldecoa_exploring_2013,aldecoa_surprise_2013,traag_significant_2013}.
In fact, numerous studies have pointed out strong theoretical limitations of
modularity approaches for community
detection~\cite{fortunato_resolution_2007,kumpula_limited_2007,lancichinetti_limits_2011,bagrow_communities_2012,xiang_limitation_2012,kehagias_bad_2012,traag_narrow_2011}.

A proposed measure based on classical probability, called
surprise~\cite{aldecoa_deciphering_2011}, has been shown to systematically
outperform modularity-based methods on different
benchmarks~\cite{aldecoa_exploring_2013,aldecoa_surprise_2013}. Here we
demonstrate how surprise can be expressed under an information-theoretic
framework, by examining its asymptotic formulation. In particular, we describe
surprise in terms of the Kullback-Leibler (KL)
divergence~\cite{kullback_information_1951}. 
This asymptotic formulation allows us to develop, for the first
time, an efficient surprise maximization algorithm. Incidentally, this
also points to a straightforward extension of surprise to weighted graphs. 
Additionally, this enables a better analysis of
its performance, and allows an analytic comparison to other methods. 

In particular, we compare surprise to a modularity model and the recently introduced
measure of significance, which also detects communities based on the
KL-divergence \cite{traag_significant_2013}. We show that surprise is more
discriminative than modularity using an Erdös-Rényi (ER) null model, and that
significance and surprise behave relatively similar.  Additionally, we analyze
the limitations of community detection, most notably the resolution
limit~\cite{fortunato_resolution_2007} and the detectability
threshold~\cite{decelle_inference_2011}.  We show that surprise is (nearly)
unaffected by the resolution limit, and works well in the limit of large number
of communities with fixed community sizes.  However, in the limit of large
community sizes with a fixed number of communities, surprise works worse than
ER modularity, as it tends to find smaller subgraphs within those larger
communities.

Apart from the choice of the null model, a key component in community detection
is how the difference between the actual community structure and the null model
is quantified. Relying on the KL-divergence to measure such difference results in
more discriminative methods. We believe that this fact can improve current and
future community detection strategies.

\section{Surprise}

In general, we denote a graph by $G = (V,E)$ consisting of nodes $V =
\{1,\ldots,n\}$ and edges $E \subseteq V \times V$, which has $n = |V|$ nodes
and $m = |E|$ links. The total number of possible links is denoted by $M =
\binom{n}{2}$, and the ratio of present links $p = \frac{m}{M}$ is known as the
density of the graph.

The general aim is to find a good partition $\mathcal{V}
=\{V_1,V_2,\ldots,V_r\}$ of the graph, where each $V_c \subseteq V$ is a set of
nodes, which we call a community. Such communities are non-overlapping (i.e.
$V_c \cap V_d = \emptyset$ for all $c \neq d$) and cover all the nodes (i.e.
$\bigcup V_c= V$). Each community consists of $n_c = |V_c|$ nodes and contains
$m_c = |E_c|$edges.  Obviously then $\sum_c n_c = n$, but the total number of
internal edges $m_{\text{int}} = \sum_c m_c$ is smaller than the total number of
edges so that $m_{\text{int}} \leq m$. An overview of the relevant variables is
provided in Table~\ref{tab:variables}.

\begin{table}[bt]
\renewcommand\arraystretch{1.15}
  \begin{tabular}{ll}
    \multicolumn{2}{l}{Graph variables} \\
    \colrule
    $n$ & Number of nodes \\
    $m$ & Number of edges \\
    $M=\binom{n}{2}$ & Number of possible edges \\
    $p=\frac{m}{M}$ & Density \\[5mm]

    \multicolumn{2}{l}{Community variables} \\
    \colrule
    $n_c$ & Number of nodes in community $c$ \\
    $m_c$ & Number of edges in community $c$ \\
    $\langle m_c \rangle$ & Expected number of edges in community $c$ \\
    $p_c=\frac{m_c}{\binom{n_c}{2}}$ & Density of community $c$ \\[5mm]

    \multicolumn{2}{l}{Partition variables} \\
    \colrule
    $m_{\text{int}} = \sum_c m_c$ & Total internal edges  \\
    $M_{\text{int}} = \sum_c \binom{n_c}{2}$ & Total possible internal edges \\ 
    $q = \frac{m_{\text{int}}}{m}$ & Fraction of internal edges \\
    $\langle q \rangle = \frac{M_{\text{int}}}{M}$ & Expected fraction of internal edges \\
  \end{tabular}
  \caption{Variables.}
  \label{tab:variables}
\end{table}

Surprise is a statistical approach to assess the quality of a partition into
communities. Given a graph with $n$ nodes, there are $M = \binom{n}{2}$ possible
ways of drawing $m$ edges. Out of those, there are $M_\text{int} = \sum_c
\binom{n_c}{2}$ possible ways of drawing an internal edge. Surprise is then
defined as the (minus logarithm of the) probability of observing at least $m_\text{int}$ successes
(internal edges) in $m$ draws \textit{without} replacement from a finite 
population of size $M$ containing exactly $M_\text{int}$ possible successes~\cite{arnau_iterative_2005,aldecoa_deciphering_2011}:
\begin{equation}
  \label{eq:original_surprise}
  \Sur(\V) = -\log \sum_{i = m_\text{int}}^{\min(m,M_\text{int})} 
    \frac{\displaystyle 
      \binom{M_\text{int}}{i} \binom{M - M_\text{int}}{m - i}}
         {\displaystyle {\binom{M}{m}}},
\end{equation}
which derives from the hypergeometric distribution.

\subsubsection{Asymptotic formulation}

However, this formulation presents some difficulties. 
It is not straightforward to work with, nor is it simple to implement in an optimization procedure, mainly due to numerical computational problems. 
Since we are usually interested in relatively large graphs, an asymptotic approximation may provide a good alternative.
The asymptotic expansion we consider here assumes that the graph
grows, but that the relative number of internal edges $q =
\frac{m_\text{int}}{m}$ and the relative number of expected internal edges
$\langle q \rangle = \frac{M_\text{int}}{M}$ remains fixed. By only considering
the dominant term, we obtain a simple and elegant approximation
(see Appendix~\ref{sec:asymptotic_analysis})
\begin{equation}
  \Sur(\V) \approx m D(q \midd \langle q \rangle),
  \label{equ:surprise}
\end{equation}
where $D(x \midd y)$ is the KL divergence
\begin{equation}
  D(x \midd y) =  x \log \frac{x}{y} + (1 - x) \log \frac{1 - x}{1 - y}.
  \label{equ:KL}
\end{equation}
The KL divergence measures the distance between two probability distributions
(although it is not a proper metric), with in this case the Bernoulli probability distributions $x$, $1-x$ and $y$, $1-y$.
Notice that, in general, $D(x \midd y) \neq D(y \midd x)$. 
In this case, $q$ and $\langle q \rangle$ denote the probability that a link lies (or is expected to lie) within a community.
Whenever $q = \langle q \rangle$, we have
that $D(q \midd \langle q \rangle) = 0$ and, otherwise, $D(q \midd \langle q
\rangle) > 0$. Since we are looking for relatively dense communities, we
generally have $q > \langle q \rangle$. 

The original formulation of surprise in Eq.~(\ref{eq:original_surprise}), based
on a hypergeometric distribution, can be accurately approximated by a binomial
distribution. The only difference between both approaches is that in the former
links are drawn without replacement.
Consider again $q =
\frac{m_\text{int}}{m}$, the fraction of internal edges in the partition, and
$\langle q \rangle = \frac{M_\text{int}}{M}$, the expected fraction of internal
edges. The binomial formulation of surprise would then be
\begin{equation}
  \Sur(\V) =  -\log \sum_{i=m_\text{int}}^{\min(m, M_\text{int})} \binom{m}{i} \,
  \langle q \rangle^{i} \, (1-\langle q \rangle)^{m-i}
  \label{equ:surprise_binomial}
\end{equation}
The asymptotic development for the dominant term of binomial surprise is
simpler. We use Stirling's approximation,
\begin{equation}
  \displaystyle \log {\binom{n}{k}} \approx n \, H\left(\frac{k}{n}\right) 
\end{equation}
where $H(x) = -x \log x - (1-x) \log (1-x)$ is the (binary) entropy and we use that $m_\text{int} = qm$. 
Binomial surprise then becomes
\begin{align*}
  \label{eq:surKL}
  \Sur(\V) &\approx - m \Bigl[H(q) + q \log \langle q \rangle  + (1 - q) \log (1 -
    \langle q \rangle) \Bigr] \\
  &= m D(q \midd \langle q \rangle)
\end{align*}
Thus, as expected, for large sparse networks the difference between drawing
with or without replacement is negligible.

\subsubsection{Algorithm}

Evaluating the quality of a partition using surprise shows excellent results in
standard benchmarks. 
In fact, it has been shown that a meta-algorithm of selecting the partition with the highest surprise, from a set of candidate solutions provided by the best community detection algorithm solutions, outperforms any single algorithm by itself~\cite{aldecoa_exploring_2013,aldecoa_surprise_2013,aldecoa_surpriseme_2014}.
However, no algorithm for directly optimizing surprise has been developed yet.

The asymptotic formulation allows a  straightforward algorithmic implementation,
in a similar fashion as the Louvain algorithm~\cite{blondel_fast_2008}, which was
initially designed to optimize modularity. The basic idea of the Louvain algorithm
consists of two steps. We move around nodes from one community to another so
as to greedily improve surprise. When surprise can no longer be improved by
moving around individual nodes, we aggregate the graph, and repeat the
procedure on the aggregated graph.

The aggregation of the graph is simply the contraction of all nodes within a
community to a single ``community node''. The multiplicities of the edges are
kept as weighted edges, so that $w_{cd} = \sum_{i \in V_c, j \in V_d} w_{ij}$
denotes the weight between the new nodes $c$ and $d$ in the aggregate graph,
where initially $w_{ij} = A_{ij}$.
Here, $A_{ij} = 1$ if there is an edge between $i$ and $j$, and $0$ otherwise.
We additionally need a node size to keep track of the total size of the
communities, similar to~\cite{traag_narrow_2011}. Initially we set this node
size to $n_i = 1$, and upon aggregation the node size $n_c = \sum_{i \in V_c}
n_i$ is set to the total number of nodes within the community.

One of the essential elements of the Louvain algorithm is that the surprise of the partition on the aggregated graph is the same as the surprise of the original partition on the original graph.
This ensures that moving a node in the aggregated graph corresponds to moving a whole community in the original graph.
In other words, if $\V$ denotes the partition of $G$ and $\V' = \{1, 2, \ldots, r\}$ denotes the default partition of the aggregated graph $G'$, then $\Sur(\V, G) = \Sur(\V', G')$.
For calculating surprise in the aggregated graph, we then use $m_c = \sum_{i,j \in V'_c} w'_{ij}$ as the internal weight and $n_c = \sum_{i \in V'_c} n_i$ as the community size and $n = \sum_c n_c$.
With the other definitions remaining the same, it is straightforward to see that $\Sur(\V, G) = \Sur(\V', G')$.
Notice that the same formulations can also be applied to the original graph, when using $w_{ij} = A_{ij}$ and $n_i = 1$.

Using this formulation of the aggregate graph, it is quite straightforward to
calculate the improvement in surprise when moving a node. Before we move node
$i$ from community $c$ to community $d$, assume we have $m_\text{int}$ internal
edges, and $M_\text{int}$ possible internal edges. The total weight between node
$i$ and community $c$ is $w_{ic} = \sum_{j \neq i \in V_c} w_{ij}$ and similarly
between node $i$ and community $d$, with a possible self-loop of $w_{ii}$. The
new internal weight after moving node $i$ from community $c$ to community $d$ is then
$m'_\text{int} = m_\text{int} - w_{ic} + w_{id}$. The change in $M_\text{int} = \sum_c \binom{n_c}{2}$
is slightly more complicated.  After the move, we obtain $n'_c = n_c - n_i$ and
$n'_d = n_d + n_i$, so that $M'_\text{int} = M_\text{int} + n_i(n_i + n_d -
n_c)$. Finally, we use $q' = \frac{m'_\text{int}}{m}$ and $\langle q' \rangle =
\frac{M'_\text{int}}{M}$. The difference in surprise for moving node $i$ from
community $c$ to community $d$ is then simply
\begin{equation}
  \Delta \Sur(\sigma_i = c \mapsto d) = 
m\left(D(q \midd \langle q \rangle) - D(q' \midd \langle q' \rangle)\right),
\end{equation}
where we denote the community of node $i$ by $\sigma_i$ (i.e. $\sigma_i = c$ if
$i \in V_c$). The algorithm can then be simply summarized as follows:

\begin{algorithmic}
  \Function{Optimizesurprise}{Graph $G$}
    \While{improvement}
      \State $\sigma_i \gets i$ for $i=1,\ldots,|V(G)|$. \Comment{Initial partition}
      \While{improvement}
        \For{random $v \in V(G)$}
          \State $\sigma_v \gets \argmax_{d} \Delta \Sur(\sigma_v = c \mapsto d)$
        \EndFor
      \EndWhile
      \State $\sigma'_i = \sigma_{\sigma'_i}$ \Comment{Community in original graph.}
      \State $G \gets$ \Call{AggregateGraph}{$G$}
    \EndWhile
    \State \Return $\sigma'$
  \EndFunction
\end{algorithmic}

Incidentally, our formulation for surprise for the aggregated graph yields a weighted version of surprise.
While keeping the same formulation of surprise as in Eq. (\ref{equ:surprise}), we only need to change the definitions of $q$ and $\langle q \rangle$. 
Then $q = \sum_c w_c / w$ where $w_c = \sum_{i,j \in V_c} w_{ij}$ is the internal weight and $w =
\sum_{ij} w_{ij}$ is the total weight. 
Assuming then a uniform distribution of weights across the graph in the random graph, the expected weights of an edge would be $\langle w \rangle$, which would not show too much deviation. 
The total possible internal weight is then $\langle w \rangle M_\text{int}$, while the total possible weight would be $\langle w \rangle M$.  
Hence, $\langle q \rangle = M_\text{int}/M$ remains unchanged.

We provide an open-source, fast and flexible \texttt{C++} implementation of the optimization of surprise using the Louvain algorithm.
It is suitable for use in \texttt{python} using the \texttt{igraph} package.
This implementation is available from \texttt{GitHub}\footnote{\url{https://github.com/vtraag/louvain-igraph}} as \texttt{louvain-igraph} and from \texttt{PyPi}\footnote{\url{https://pypi.python.org/pypi/louvain/}} simply as \texttt{louvain} and implements various other methods as well.

\section{Comparison}

We now review how surprise compares to some closely related methods. There
are many other methods still, and we cannot do all of them justice here. For a
more comprehensive review, please refer
to~\cite{fortunato_community_2010,porter_communities_2009}. 

\subsection{Modularity}

Although relatively recent, modularity has rapidly become an extremely popular
method for community detection. The general idea is that we want to find a
partition, such that the communities have more internal links than expected.
In its original formulation, modularity assumes a null model in which the
degree $k_i$ of a node is fixed~\cite{newman_finding_2004}, the so called
configuration model~\cite{molloy_critical_1995}. This implies that the expected
number of internal edges is
\begin{equation}
  \langle m_c \rangle = \frac{K_c^2}{4m},
\end{equation}
where $K_c = \sum_{i \in V_c} k_i$ is the total degree of nodes in community $c$.
Modularity compares this value to the observed number of edges $m_c$ within the 
community, and simply sums the difference. The measure is usually normalized by the
total number of edges, obtaining
\begin{equation}
  \Q_\text{CM}(\V) = \frac{1}{m} \sum_c \left(m_c - \frac{K_c^2}{4m}\right).
\end{equation}
This random graph null model represents the configuration model, where the
degree dependency of the nodes is taken into account. We therefore refer to it
as the CM modularity.

Alternative derivations of modularity have been proposed, some of them with
different null models~\cite{reichardt_statistical_2006}. 
Surprise implicitly assumes a null model in which every edge appears with the
same probability $p$, as in an ER random graph.
The number of expected edges in a community of size $c$ is thus
\begin{equation}
  \langle m_c \rangle = p \binom{n_c}{2}.
\end{equation}
Plugging this null model into modularity, we obtain its ER
version~\cite{reichardt_statistical_2006}
\begin{equation}
  \Q_\text{ER}(\V) = \frac{1}{m} \sum_c \left(m_c - p \binom{n_c}{2} \right),
\end{equation} 

There is an interesting relationship between this ER modularity and surprise.
Given that $p = m/M$, we can write
\begin{align}
  \Q_\text{ER}(\V) &= \sum_c \frac{m_c}{m} - \sum_c \frac{ \binom{n_c}{2} }{M} \\
         &= q - \langle q \rangle.
\end{align}
By Pinsker's inequality this is related to the KL divergence as 
\begin{equation}
  q - \langle q \rangle \leq \sqrt{\frac{1}{2} D(q \midd \langle q \rangle)},
\end{equation}
and, therefore,
\begin{equation}
  \Sur(\V) = mD(q \midd \langle q \rangle) \geq 2m\Q_\text{ER}(\V)^2.
  \label{equ:RBER_inequality}
\end{equation}
This implies that whenever surprise is low, modularity is also low.
Whenever a good partition (in the sense of being different from random) cannot be found by surprise, it is unlikely that modularity will be able to find one.
While Eq.~(\ref{equ:RBER_inequality}) is sometimes tight, on some partitions surprise can be much larger than modularity, making it more likely to be selected as optimal while escaping the scrutiny of modularity optimization. 
In this sense, surprise is more discriminative than modularity

\begin{figure}[t]
  \begin{center}
    \includegraphics{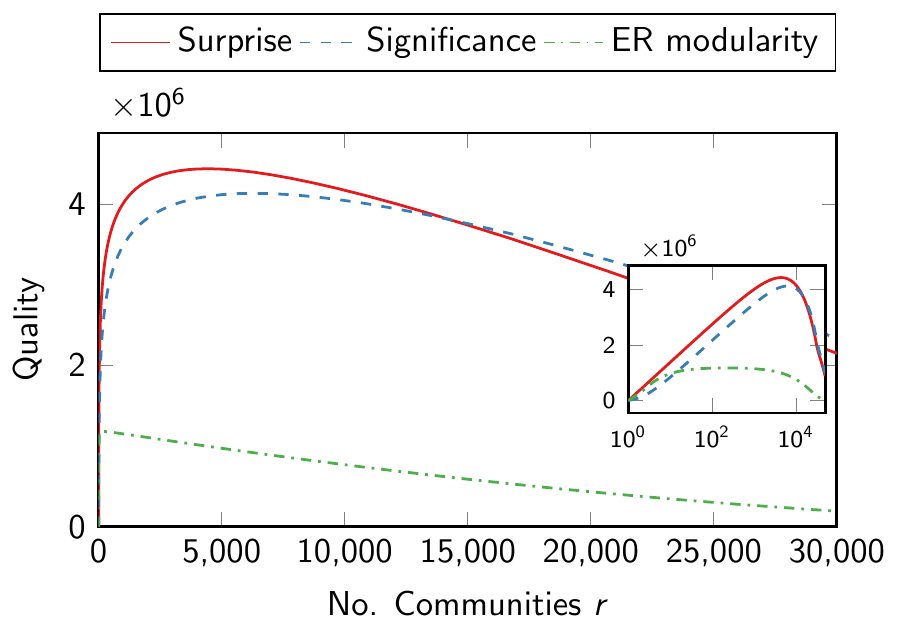}
  \end{center}
  \caption{(Color online) Comparison of bounds. 
    We show the quality of partitions of a lattice in $r$ communities.
    ER modularity quickly reaches a maximum for few communities (we show $2mQ_\text{ER}^2$ rather than $Q_\text{ER}$ for comparison).
    Both significance and surprise reach a maximum for much more communities. 
    This illustrates that ER modularity is simply unable to discern partitions with such a high number of communities.
    The inset shows the same data, but on a logarithmic x-axis.
  }
  \label{fig:lattice}
\end{figure}

To illustrate this, consider a one dimensional circular lattice with neighbors within distance $3$. 
In other words, node $i$ is connected to nodes $i - 3 \mod n$ to $i + 3 \mod n$ (excluding the self-loop).
We create partitions consisting of $r$ communities by grouping consecutive nodes such that $n/r$ nodes are in the same community. 
The ER modularity reaches its maximum with just a few communities (Fig.~\ref{fig:lattice}). 
Modularity indeed often detects only few communities, part of the problem of its resolution limit~\cite{fortunato_resolution_2007,kumpula_limited_2007,traag_narrow_2011}.
Both surprise and significance (see next section), still increase whereas ER modularity is already decreasing again.
ER modularity may not be able to discern partitions with many communities, whereas surprise and significance can.
On the other hand, when surprise goes to $0$ we see that ER modularity indeed also goes to $0$, showing the upper bound provided by surprise.

\subsection{Significance}

Significance~\cite{traag_significant_2013}, a recently introduced objective function to
evaluate community structure quality, presents an approach similar to surprise.
Surprise describes how likely it is to observe $m_\text{int}$ internal links in communities.
Significance, on the other hand, looks at how likely such dense communities appear in a random graph. 
Comparing the two measures is not immediately straightforward.
On the one hand, if dense communities are unlikely to be present in a random graph (high significance), then a community is also unlikely to contain many links at random (high surprise).
On the other hand, if a community is unlikely to contain many links at random (high surprise), perhaps there are still communities elsewhere in the random graph that contain so many links.
Therefore we should compare the two more formally to make more exact statements.

Asymptotically, significance is defined as
\begin{equation}
  \Sig(\V) = \sum_{c} \binom{n_c}{2} D( p_{c} \midd p),
\end{equation}
where $p_{c} = m_{c} / \binom{n_c}{2}$ is the density of community $c$, $p$
is the density of the graph and $D(x \midd y)$ is again the KL
divergence. Significance also showed a great performance in standard
benchmarks, and helped to determine the proper scale of resolution in
multi-resolution methods~\cite{traag_significant_2013}.

Both surprise and significance are based on the KL divergence to compare the
actual number of internal edges to the expected one. However, they do so in
different ways. Whereas surprise compares such difference using global
quantities, $q$ and $\langle q \rangle$, significance compares each community
density $p_c$ to the average graph density $p$.

This implies, among other things, that only significance is affected by the
actual distribution of edges between communities. In particular, moving edges
from a denser community (with a high $p_c$) to a sparse community (with a low
$p_c$), generally decreases the value of significance. This means that if all
communities have the same density, \emph{ceteris paribus}, significance is
minimal. This intuition is confirmed by convexity of the KL divergence (see
Appendix~\ref{sec:significance}), so that significance is lower-bounded by 
\begin{equation}
  \Sig(\V) \geq M_\text{int} D( \langle p_c \rangle \midd p)
\end{equation}
with the weighted average density
\begin{equation}
  \langle p_c \rangle = \sum_c \frac{ \binom{n_c}{2} }{M_\text{int}} p_c = \frac{m_\text{int}}{M_\text{int}} = p \frac{q}{\langle q \rangle}.
\end{equation}
Convexity of the KL divergence, also shows that
\begin{align}
    \Sig(\V) &\geq \Sur(\V)
\end{align}
whenever $\langle q \rangle < p$ (see Appendix~\ref{sec:significance}). To gain
more insight, we can slightly rewrite $\langle q \rangle$ to obtain 
\begin{equation}
  \langle q \rangle = \frac{\sum_c \binom{n_c}{2}}{\binom{n}{2}} 
                    \approx \frac{\sum_c n_c^2}{n^2} 
                    = \frac{1}{r} \frac{\langle n_c^2 \rangle}{\langle n_c \rangle^2}.
\end{equation}
Then, in general, $\langle q \rangle$ will be inversely proportional to the
number of communities, and increases with the variance of the community sizes
$n_c$. Hence, if the number of communities is relatively large (small $\langle q
\rangle$), or the network is relatively dense (large $p$), significance is more
discriminative than surprise. However, in the case that $\langle q \rangle > p$,
surprise can be more discriminative than significance (see
appendix~\ref{sec:significance}). Notice that if $\langle q \rangle = p$, then
$p_c = q$, so that $D(\langle p_c \rangle \midd p) = D(q \midd \langle q \rangle)$ 
and significance and surprise values are close to each other.  Therefore, the two
measures are expected to behave relatively similar, especially for $\langle q
\rangle \approx p$. Nonetheless, in dense networks with many communities significance 
would be more discriminative, whereas for fewer communities or sparse graphs,
surprise would show a better performance.

\section{Limitations}

Although modularity was lauded by the possibility to detect communities without specifying the number of communities, this came at a certain price.
One of the best known problems in community detection is the resolution limit~\cite{fortunato_resolution_2007}, which prevents modularity from detecting small communities.
It thus tends to underestimate the number of communities in a graph, and lumps together several smaller communities in larger communities.
Moreover, this depends on the scale of the graph, so that modularity has a problem of scale.
It was shown that this is the case for both ER and CM modularity, and that other null models also suffer from the same drawbacks~\cite{kumpula_limited_2007}.
In fact, most methods are expected to suffer from this problem, and only few methods are able to avoid it completely~\cite{traag_narrow_2011}.
Additionally, there is also a lower counterpart to the resolution limit, leading to unnecessary splitting of cliques~\cite{krings_upper_2011,traag_algorithms_2014}.
Finally, modularity is also myopic, cutting across long dendrites~\cite{schaub_markov_2012}.
Another fundamental limit in community detection is called the detectability threshold~\cite{decelle_inference_2011}, which also has some counter-intuitive effects~\cite{radicchi_paradox_2014}.
This prevents any method from correctly detecting communities beyond this threshold.
The asymptotic formulation of surprise enables us to understand better how it performs with respect to these limitations.

\subsection{Resolution limit}

The resolution limit is traditionally studied through the ring of cliques~\cite{fortunato_resolution_2007}.
This is a graph consisting of $r$ cliques (i.e. completely connected subgraphs) connected only by one link between two cliques to form a ring.
This is one of the most modular structure possible: we cannot delete more than one link between communities and still keep it connected, while we cannot add any more links within the cliques.
When a method starts to join the cliques, it can no longer detect the smaller cliques, and so \emph{a fortiori}, cannot detect less well defined subgraphs either.
We denote by $q_1$ (and $\langle q_1 \rangle$) the (expected) proportion of edges within communities for the partition where each community contains a single clique and use $q_2$ (and $\langle q_2 \rangle$) for the partition where each community contains two cliques. 
To facilitate the derivation, we work with self-loops (and directed edges), so that the total number of edges is $n_c^2$ within communities respectively. 
Let $r$ denote the number of cliques. 
Then obviously $n = rn_c$ and $m = rn_c^2 + 2 r$.
For the partition of each clique in its own community we then obtain 
\begin{align}
  q_1 &= \frac{n_c^2}{n_c^2 + 2}, & \langle q_1 \rangle &= \frac{1}{r},
\end{align}
while for the partition with $2$ cliques merged we obtain 
\begin{align}
  q_2 &= \frac{n_c^2 + 1}{n_c^2 + 2}, & \langle q_2 \rangle &=  \frac{2}{r}.
\end{align}
Hence, $q_2 = q_1 + \epsilon$ with $\epsilon = \frac{1}{n_c^2 + 2}$ and $\langle q_2 \rangle = 2 \langle q_1 \rangle$. The difference of surprise is
\begin{equation}
  \Delta \Sur = \frac{\Sur_2 - \Sur_1}{m} = D(q_2 \midd \langle q_2 \rangle) - D(q_1 \midd
  \langle q_1 \rangle) 
\end{equation}
which works out to
\begin{multline}
  \Delta \Sur = q_1 \log \frac{q_2}{\langle q_2 \rangle} \frac{\langle q_1 \rangle}{q_1} + 
  (1 - q_1) \log \frac{1 - q_2}{1 - \langle q_2 \rangle} \frac{1 - \langle q_1 \rangle}{1 - q_1} + \\ 
  \epsilon \log \frac{q_2}{\langle q_2 \rangle} \frac{1 - \langle q_2 \rangle}{1 - q_2}.
\end{multline}
Approximating $r - 2 \approx r - 1 \approx r$ we obtain 
\begin{equation}
  \Delta \Sur \approx - D(q_1 \midd q_2) - q_1 \log 2 + \epsilon \log \frac{r}{2} \frac{q_2}{1 - q_2}.
\end{equation}
Solving for $r$ at the point at which $\Delta \Sur = 0$ yields
\begin{equation}
  r = 2 \frac{1 - q_2}{q_2} \exp \left( \frac{1}{\epsilon} D(q_1 \midd q_2) \right) 2^{\frac{q_1}{\epsilon}}
\end{equation}
which scales as $r \sim \frac{2^{n_c^2}}{n_c^2}$ so that for larger $r$ surprise starts to merge cliques.

Working out the inequality for both CM and ER modularity we obtain that $r \sim n_c^2$.
Hence, the number of cliques $r$ at which modularity starts to merge cliques lies considerably lower than for surprise and grows linearly with the square of community sizes rather than exponentially.
So, although surprise shows a similar problem as modularity, it only starts to show at really large graphs, so is unlikely to be a problem in any empirical graph.
Indeed, this demonstrates exactly the key difference between modularity and surprise: The first is unable to detect relatively small communities in large graphs, whereas the latter has (nearly) no such difficulties.

\subsection{Detectability threshold}

\begin{figure*}[t]
  \begin{center}
    \includegraphics{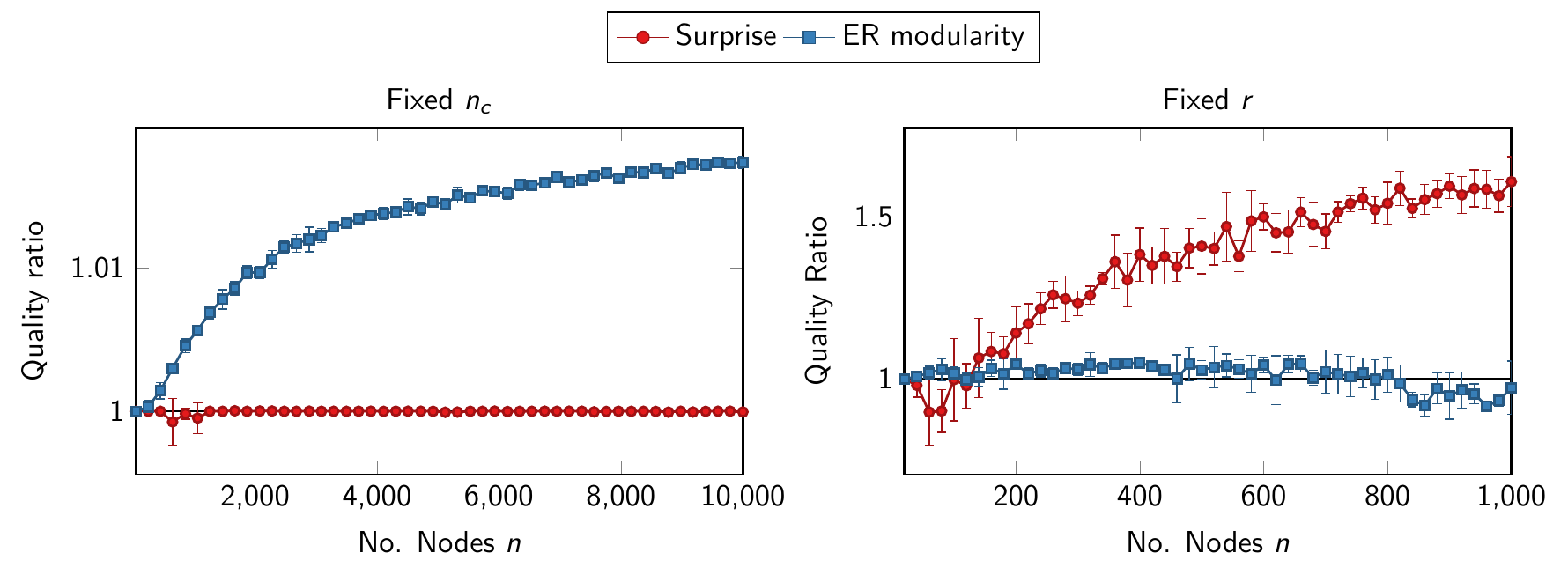}
  \end{center}
  \caption{(Color online) Limitations on community detection. 
    We construct graphs with a planted partition, with a probability of an edge between communities of $\mu = 0.1$. 
    We show the quality ratio $\frac{\Sur}{\Sur_\text{plt}}$ between the quality of the partition found by optimization $\Sur$ and the quality of the planted partition $\Sur_\text{plt}$ (and similarly for ER modularity).
    Hence, if the quality ratio $\frac{\Sur}{\Sur_\text{plt}} > 1$, the planted partition is no longer optimal.
    In the figure on the left we consider the case for fixed community size $n_c = 10$, but increase the number of communities $r$.
    The results show that in this case surprise finds the planted partitions, whereas ER modularity has more difficulties, in line with our analysis.
    This is mostly due to the resolution limit in modularity, which underestimates the number of communities.
    In the figure on the right we consider the case of a fixed number of communities $r=2$ but an increasing community size $n_c$.
    In this case, surprise quickly finds other partitions than the planted partition, whereas modularity remains closer to the planted partition, consistent with our analysis.
    This is mostly because surprise tends to find substructure in the rather large communities arising from random fluctuations, which thus overestimates the number of communities.
    However, modularity also has some difficulty in finding the planted partition.
    This figure shows the average over $5$ replications for each setting, and the error bars show the standard deviation.}
  \label{fig:limitations}
\end{figure*}

In order to study the detectability threshold, we first introduce the planted partition model.
This means, that we build a graph such that it will contain a specified partition: We plant it in the graph.
We create $n$ nodes and assign each node to a certain community.
An edge within a community is created with probability $p_\text{in}$, whereas an edge in between two communities is created with probability $p_\text{out}$.
We define the probability of an internal edge $p_\text{in}$ and the probability of an external edge to be respectively
\begin{align}
  p_\text{in} &= \frac{(1 - \mu)k}{n_c - 1}, &
  p_\text{out} &= \frac{\mu k}{n - n_c},
\end{align}
so that the average degree is $k$ and $\mu$ is the probability that an edge is between communities.
When $\mu = 0$ all links are thus placed within the planted communities, whereas for $\mu=1$ all links are placed between the planted communities.
Uncovering the planted communities correctly is trivial for $\mu = 0$ but becomes increasingly more difficult for higher $\mu$.
The average degree within a cluster is $k_\text{in} = (1 - \mu)k$ while the average degree between clusters is $k_\text{out} = \mu k$.
We denote community sizes by $n_c$ for the $r$ different communities.

Notice that, most conveniently, $q = 1 - \mu$, while $\langle q \rangle = \frac{1}{r}\frac{\langle n_c^2 \rangle}{\langle n_c \rangle^2}$.
We can thus easily calculate $\Sur_\text{plt}$ the surprise for the planted partition.
Since $\Sur > 0$ by definition, communities can thus only be detected when $1 - \mu > \frac{1}{r}\frac{\langle n_c^2 \rangle}{\langle n_c \rangle^2}$.
This yields the rather trivial detectability threshold of
\begin{equation}
  \mu < \frac{r - \frac{\langle n_c^2 \rangle}{\langle n_c \rangle^2}}{r}.
\end{equation}
In the case of equi-sized communities, this reduces to the familiar trivial threshold $\mu < \frac{r - 1}{r}$~\cite{lancichinetti_community_2009}.

However, due to stochastic fluctuation, the communities become already ill-defined prior to the threshold.
Indeed $\Sur = 0$ provides a rather naive bound, since $\Sur > 0$ also in random graphs.
In general, $\Sur = 0$ for both trivial partitions of one large community and $n$ small communities (since then $q = \langle q \rangle$), so that optimizing surprise in a random graph will yield some partition with strictly positive surprise.
This implies that at some (lower) critical $\mu^*$ the community structure is essentially no longer discernible from the community structure in a random graph.
Hence, we should not consider when $\Sur_\text{plt} > 0$ but when $\Sur_\text{plt} > \Sur_\text{rnd}$ where $\Sur_\text{rnd}$ is the surprise attainable in a random graph.
We first examine the case with $r = 2$ and $n_c = n/2$.
Previous literature found a detectability threshold for $k_\text{in} - k_\text{out} \leq \sqrt{k_\text{in} + k_\text{out}}$~\cite{decelle_inference_2011,nadakuditi_graph_2012,radicchi_detectability_2013}.
Beyond this threshold, the optimal bisection becomes indiscernible from an optimal bisection in a random graph.
This threshold thus coincides with the expected number of internal edges for an optimal bisection in a random graph.
We can use this to calculate $\Sur_\text{rnd}(2)$ the maximum surprise for a bisection in a random graph.
Let us denote by $q_\text{rnd}(2)$ the probability an edge is within a community in the best bisection for a random graph.
Substituting $k_\text{in} = q_\text{rnd}(2) k$ and $k_\text{out} = (1 - q_\text{rnd}(2)) k$ and solving for $q_\text{rnd}(2)$ yields 
\begin{equation}
  q_\text{rnd}(2) = \frac{1}{2}\left(1 + \sqrt{\frac{1}{k}}\right).
  \label{equ:detectability}
\end{equation}
We thus obtain $\Sur_\text{rnd}(2) = m D(q_\text{rnd}(2) \midd \frac{1}{2})$ for the maximum surprise for a bisection in a random graph.
If $\Sur_\text{plt}(2) < \Sur_\text{rnd}(2)$ the planted partition is no longer optimal, and we will likely find an alternative partition with surprise equal to $\Sur_\text{rnd}(2)$.
The threshold is then $\mu^* = 1 - q_\text{rnd}(2)$, congruent with previous results.
So, in general, surprise is expected to show similar behavior concerning the detectability threshold as other methods.

However, this analysis restricts itself to finding the same number of communities (i.e. two in this case), while it is possible that an optimal partition would split the graph in more communities.
In other words, we need to compare the surprise of the planted partition to the maximum surprise in a random graph, while allowing more than two communities.
Although the expected value of the maximum surprise in a random graph is not easy to find, a random graph is likely to contain a near perfect matching.
Using that, we can derive a lower bound on the expected surprise in a random graph. 
In such a perfect matching there are $r = \frac{n}{2}$ communities which contain $1$ link each.
For a graph that contains $m = n k$ edges, then $q = \frac{1}{2k}$ while $\langle q \rangle = \frac{2}{n}$.
This leads to a surprise of approximately $\Sur_\text{rnd}(\frac{n}{2}) \sim \frac{n}{2} \log \frac{n}{4k}$.
Hence, whenever we obtain that $\Sur_\text{plt} \leq \Sur_\text{rnd}(\frac{n}{2})$ optimization should find another partition than the planted one.
In the case of two planted communities, we require that $D( 1 - \mu \midd \frac{1}{2}) \geq \frac{\log \frac{n}{4 k}}{ 2 k }$ to make sure that we still detect the two clusters.
Although we cannot solve explicitly for $\mu$, this inequality shows that $n$ is bounded above by
\begin{equation}
  n \leq 4 k e^{2 k D\left(1 - \mu \midd \frac{1}{2}\right)}.
\end{equation}
If $n$ grows large, there is likely some structure arising from random fluctuations within the planted communities. 
Notice that there are likely better partitions than a perfect matching.
We can therefore expect the actual critical $n$ for which the planted partition is no longer optimal to be lower.

We can similarly derive such thresholds for ER modularity.
For a perfect matching the ER modularity is $\mathcal{Q}_\text{rnd}(\frac{n}{2}) = \frac{1}{2k} - \frac{2}{n}$.
Then solving $\mathcal{Q}_\text{plt} \leq \mathcal{Q}_\text{rnd}(\frac{n}{2})$ gives us an estimate of when ER modularity is likely to find an alternative partition (i.e. a perfect matching in this case).
The critical $\mu^*$ can in this case be explicitly derived and yields $\mu^* = \frac{1}{2}\left(1 - \frac{1}{k} + \frac{4}{n}\right)$.
However, the detectability threshold is already reached before that point at $\mu^* = \frac{1}{2}\left(1 - \sqrt{\frac{1}{k}}\right)$, leaving $n$ essentially unbounded.
Again, there will be better partitions than a perfect matching, so that $n$ may still be bounded to some extent.
Nonetheless, this shows that ER modularity is less affected by the size of the communities than surprise, and is less likely to find substructure within the planted communities.

In summary then, surprise does not tend to suffer from the resolution limit, but does quickly find substructure due to random fluctuations.
ER modularity on the other hand suffers from a resolution limit, but tends to ignore substructure in communities.
Stated differently, for a planted partition model with $r$ communities and $n = r n_c$ nodes, surprise and ER modularity work well in different limits.
Whenever $r \to \infty$ with $n_c$ fixed, surprise works well but ER modularity works poorly.
Whenever $r$ is fixed but $n_c \to \infty$, ER modularity works well, but surprise works poorly.
An interesting question would concern which method would work well for both limits.

\section{Experimental Results}

We here confirm our theoretical results experimentally. 
We first show numerically that the asymptotic formulation of surprise provides an excellent approximation. 
Secondly, we validate the inequalities between surprise, significance and ER modularity. 
Thirdly, we show the different limitations on surprise and modularity. 
Finally, we demonstrate that the asymptotic formulation of surprise performs very well in LFR benchmarks~\cite{lancichinetti_benchmark_2008}.

\begin{figure}[t]
  \begin{center}
    \includegraphics[width=\columnwidth]{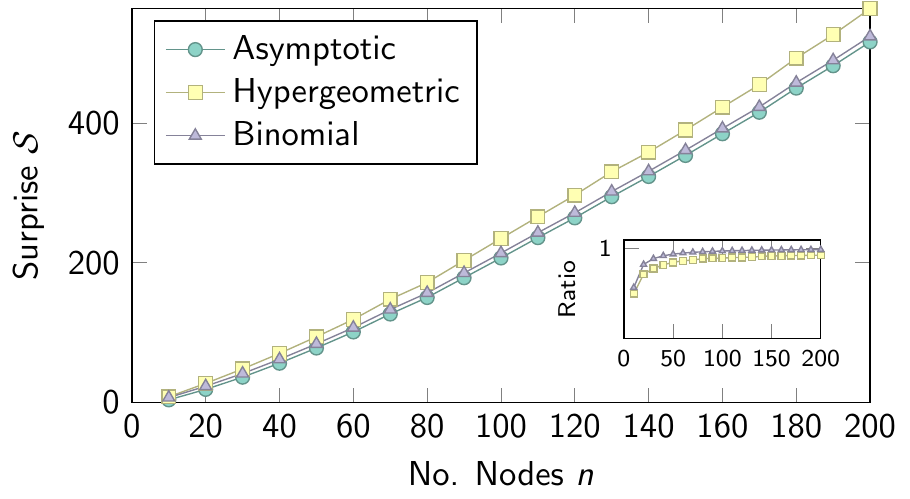}
  \end{center}
  \caption{
    (Color online) Approximation of surprise. The asymptotic formulation of surprise, using
    the KL divergence, approximates well both the binomial and the hypergeometric
    surprise. The inset shows the approximation ratio
    $\Sur_\text{asym}/\Sur_\text{hyper}$ and $\Sur_\text{asym}/\Sur_\text{binom}$,
    both going to $1$ for large graphs.
  \label{fig:approximation}}
\end{figure}

For comparing the asymptotic formulation with the exact hypergeometrical and binomial formulation, we used regular rooted trees with three children.
To create such trees, we first create the root node, and add three children to this root node. 
We then keep on adding children to the leaves of the tree until we obtain the desired number of nodes.
We use trees to minimize the number of edges to prevent numerical problems with the hypergeometrical and binomial formulation. 
Using relatively large numbers results in numerical issues, preventing a comparison to the asymptotic formulation. 
We optimize asymptotic surprise using the Louvain algorithm to find a partition on this graph.
As can be seen in Fig.~\ref{fig:approximation}, the approximation is quite
good, and the approximation ratio tends to $1$. 
Notice that the number of nodes in these graphs is limited to $200$, whereas complex networks are usually much larger. Hence, we expect the approximation to be accurate for any real network.

To demonstrate the limitations on surprise and (ER) modularity we create some test networks with a planted partition.
We generate networks with average degree $\langle k \rangle = 10$ and set $\mu = 0.1$.
In the first test, we create networks with fixed community sizes $n_c = 10$ and vary the number of communities $r$.
In the second test, we have fixed the number of communities to $2$ but vary the community size $n_c$ from $10$ to $500$.
We consider whether the planted partition remains optimal by analyzing the quality of the planted partition $\Sur_\text{plt}$ (or $\Q_\text{plt}$ for modularity) and the partition found through optimization $\Sur$ (or $\Q$ for modularity).
Whenever $\Sur_\text{plt} < \Sur$ we thus know that the planted partition remains no longer optimal.
The results shown in Fig.~\ref{fig:limitations} clearly confirm our theoretical analysis.
In the case where $r \to \infty$ with fixed $n_c$, surprise does well, whereas (ER) modularity suffers from the resolution limit.
In the case that $r$ is fixed to $r=2$, but $n_c \to \infty$, surprise does less well, as it tends to find subgraphs within the two large communities.
Modularity also has problems identifying the optimal bisection.
Indeed, the uncovered partitions do not coincide exactly with the planted partition, even though the modularity value remains rather similar.
Such partitions are likely to occur because of the degeneracy of modularity~\cite{good_performance_2010}.
Nonetheless, our results show that the modularity of the planted partition remains (nearly) optimal, whereas surprise for the planted partition clearly diminishes compared to surprise of the uncovered partitions.

We also tested the various methods more extensively using benchmark graphs with a more realistic community size and degree distribution~\cite{lancichinetti_benchmark_2008}.
We set the average degree $\langle k
\rangle = 20$ while the maximum degree is $50$ and follows a powerlaw degree
distribution with exponent $2$. Planted community sizes range from $10$ to $50$
for the ``small'' communities, and from $20$ to $100$ for ``large'' communities.
The planted community sizes are also distributed according to a powerlaw, but with an
exponent of $1$. 
The parameter $\mu$ again controls the probability of internal links.

\begin{figure}[t]
  \begin{center}
    \includegraphics[width=\columnwidth]{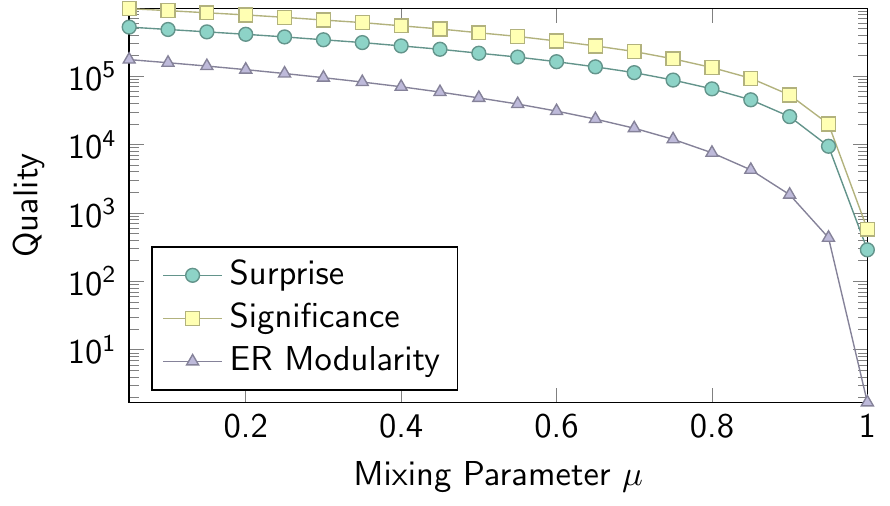}
  \end{center}
  \caption{(Color online) Inequalities. 
    In most cases significance is more discriminative than surprise, which is more discriminative than the ER modularity, so that $\Sig > \Sur > \Q_\text{ER}$. 
    These inequalities clearly hold over the whole range of the mixing parameter $\mu$ for LFR benchmarks ($n = 10^4$).
    For ER modularity we display $2 m \Q_\text{ER}^2$ as used in Eq. (\ref{equ:RBER_inequality}).
  }
  \label{fig:inequalities}
\end{figure}

In Fig.~\ref{fig:inequalities} we show the function values for surprise,
significance and ER modularity. This clearly shows that the inequalities hold
over the whole range of mixing parameters. At the same time, they show very
similar behavior to each other. Although this could indicate a relatively
similar performance, we next show this is not the case.

\begin{figure*}[t]
  \begin{center}
    \includegraphics[width=\textwidth]{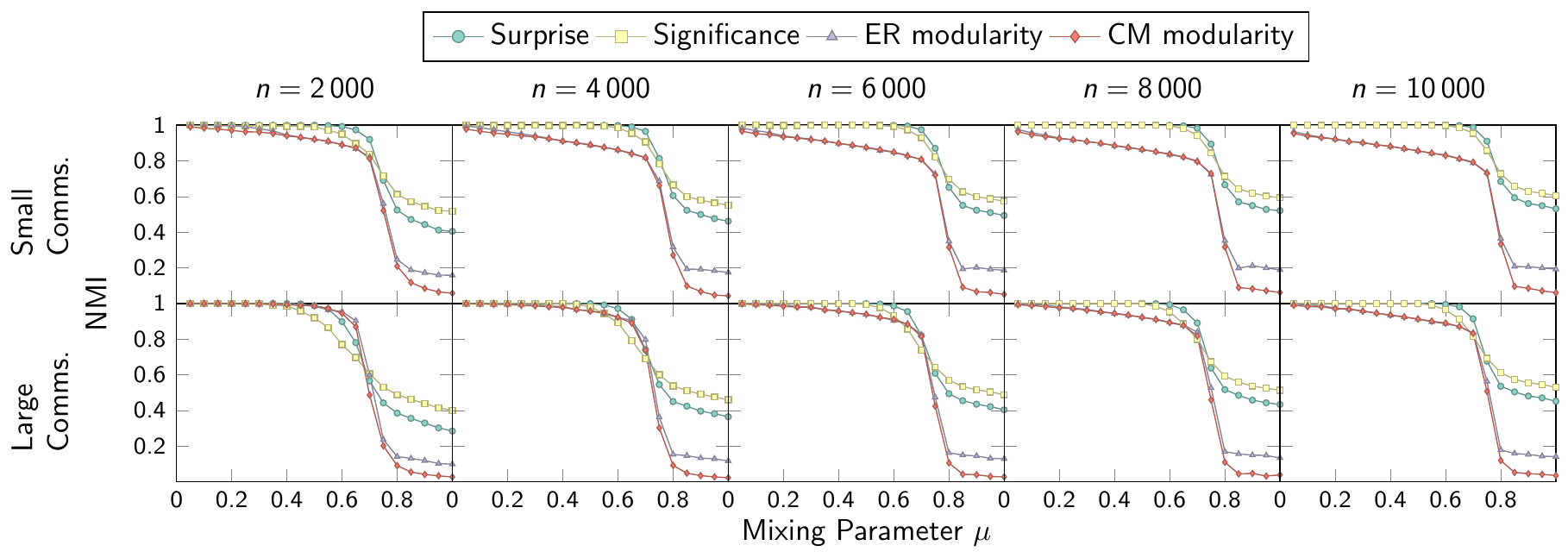}
  \end{center}
  \caption{(Color online) Benchmark results. 
    The first row shows results for ``small'' communities, which range from $10$--$50$, while the second row contains results for ``large'' communities, ranging from $20$--$100$. 
    The community sizes are powerlaw distributed with exponent $1$.
    We set the average degree $\langle k \rangle = 20$ and the maximum degree is $50$, which follows a powerlaw degree distribution with exponent $2$.
    Both surprise and significance perform very well, especially for relatively large
    graphs, where ER and CM modularity fail. This difference is more
    notable for smaller communities, for which both ER and CM modularity have
    difficulties. This is in part due to the well-known resolution limit and in line with our earlier analysis.
  }
  \label{fig:benchmark}
\end{figure*}

In Fig.~\ref{fig:benchmark} we show the benchmark results for the four different methods.
Surprise and significance performances are very good, and clearly much better than both modularity models.
Notice that, surprise and ER modularity use the same global quantities. 
However, the use of the KL divergence gives the former a much greater advantage, as expected from Eq.~(\ref{equ:RBER_inequality}).

LFR benchmark graphs have a clearer community structure for larger graphs. The
critical mixing parameter at which the inner community density equals the outer
community density is roughly $\mu \approx 1 - \frac{n_c}{n}$, so that with
growing $n$ this threshold goes to $1$. Both surprise and significance start to
work better for somewhat larger graphs, consistent with the clearer community
structure. This is in a sense the opposite of both ER and CM modularity.
Their performance is worse for larger graphs, consistent with our earlier analysis of the limitations of community detection.

\section{Conclusion}

Community detection is an important topic in the field of complex networks, as it can give us a better understanding of real-world networks.
Here we analyzed a recent measure known as surprise. 
We developed an accurate asymptotic approximation, based on the KL divergence which we use to develop a competitive new algorithm. 
Applying this algorithm to standard benchmarks, we show its great potential. 
Significance, another quality measure also based on the KL divergence performs similar to surprise.

We showed analytically that surprise is more discriminative than modularity with
an ER null model. 
This is mainly due to the use of the KL divergence to quantify the difference between the empirical partition and the null model. 
The larger the network and the smaller the communities, the better KL methods perform with respect to modularity. 
Indeed, whereas modularity suffers from the resolution limit, this problems (nearly) doesn't affect surprise.
On the other hand, surprise tends to find substructure in larger communities, arising from random fluctuations, whereas this problems appears less prominent for modularity.
In short, modularity tends to work well in the limit of community sizes $n_c \to \infty$ keeping the number of communities $r$ fixed.
Surprise on the other hand works well when $r \to \infty$ keeping the community sizes $n_c$ fixed.
Stated differently, modularity tends to underestimate the number of communities, whereas surprise tends to overestimate the number of communities.
The question of which method works well in both limits deserves further study.

The slight differences between surprise and significance stem from two things either the one or the other measure ignores. 
Significance relies on the fraction of edges that are present within a community. 
It thus implicitly considers missing edges within communities, because this fraction is relative to the total number of possible edges within that community, which surprise does not.
Surprise on the other hand, considers the fraction of total edges that fall within communities.
It thus implicitly considers edges that fall between communities, whereas significance does not.
Indeed, it should be possible to address these shortcomings by also explicitly
examining missing links (for surprise) or links between communities (for
significance).

Another shortcoming is that surprise does not depend on the actual distribution of the internal edges among communities.
One way to address this issue is to consider edges for all communities separately, by using a multivariate hypergeometric distribution.
In that case, we would be interested in the probability to observe $m_{cd}$ edges between communities $c$ and $d$ as
\begin{equation}
  \label{equ:surprise_blockmodel}
  \Pr(X_{cd} = m_{cd}) = \frac{ 
    \displaystyle \prod_{cd} \binom{n_c n_d}{m_{cd}} }{
    \displaystyle \binom{M}{m} }.
\end{equation}
Again deriving an asymptotic expression, we arrive at
\begin{equation}
  \Sur(\V) = m \sum_{cd} q_{cd} \log \frac{q_{cd}}{ \langle q_{cd} \rangle } 
  = m D(\mathbf{q} \midd \langle \mathbf{q} \rangle)
\end{equation}
where $q_{cd} = \frac{m_{cd}}{m}$ is the fraction of edges between communities $c$ and $d$ and $\langle q_{cd} \rangle$ the expected value.

Interestingly, the extension of surprise in Eq.~(\ref{equ:surprise_blockmodel})
is identical to a stochastic blockmodel (using an ER null
model)~\cite{karrer_stochastic_2011,bickel_nonparametric_2009}.  However, Karrer
and Newman found that this method did not work
satisfyingly~\cite{karrer_stochastic_2011}. This might be because the measure
does not focus on communities specifically, but rather on all types of block
structures. Hence, there is no reason why a community structure should maximize
this likelihood, rather than any other type of block structure. One possible way
to address this is to compare our partition to the ideal type we are looking
for, rather than maximizing the difference to a random null model. This would be
an interesting avenue to consider in future research.

\bibliography{mybib}

\appendix

\section{Asymptotic surprise}
\label{sec:asymptotic_analysis}

As stated in the main text, $q$ denotes the fraction of internal edges, so that
we can write $m_\text{int} = q m$. Since $m = p \binom{n}{2} = p M$, we thus
have $m_\text{int} = q p M$. Similarly, we can write $M_\text{int} =
\langle q \rangle M$. Hence, we obtain
\begin{align}
  m &= p M, \\
  m_\text{int} &= q p M, \\
  M_\text{int} &= \langle q \rangle M.
\end{align}
Notice that all quantities now depend on $M$. We only take into account the
dominant term, so to obtain
\begin{equation}
  \Sur(\V) \approx -\log \frac{ \displaystyle 
    \binom{\langle q \rangle M}{p q M} 
    \binom{(1 - \langle q \rangle) M}{p (1 - q) M} }{ \displaystyle
      \binom{M}{p M}  }
\end{equation}
which corresponds to the probability of observing exactly $m_\text{int}$
internal links. The binomial coefficient $\binom{M}{pM}$ is independent of the
partition, so we ignore it. We use Stirling's approximation of the binomial
coefficient which reads
\begin{equation}
  {\binom{n}{k}} \approx \left(\frac{n}{k}\right)^{k}.
\end{equation}
Hence, for the dominant term, we obtain
\begin{align}
  \Sur(\V) &= -\log \left(\frac{\langle q \rangle M}{p q M}\right)^{pqM} 
                    \left(\frac{(1 - \langle q \rangle)M}{p(1 - q)M}\right)^{p(1-q)M} \\
           &= -\log p^{-pN}
           \left( \left(\frac{\langle q \rangle}{q}\right)^q\left(\frac{1 - \langle q \rangle}{1-q}\right)^{1-q}
           \right)^{pM}.
\end{align}
The term $p^{-pM}$ is independent of the partition and we ignore it, which yields 
\begin{equation}
  \Sur(\V) = - pM \left( q \log \frac{\langle q \rangle}{q} + (1 - q) \log \frac{1 - \langle q \rangle}{1 - q} \right).
\end{equation}
Using $pM = m$, we can rewrite this to 
\begin{equation}
  \Sur(\V) = mD(q \midd \langle q \rangle)
  \label{equ:asymp_hypergeometric}
\end{equation}
where $D(x \midd y)$ is the KL
divergence~\cite{kullback_information_1951}
\begin{equation}
  D(x \midd y) =  x \log \frac{x}{y} + (1 - x) \log \frac{1 - x}{1 - y},
\end{equation}
which can be interpreted as the distance between the two probability
distributions $q$ and $\langle q \rangle$.

\section{Significance}
\label{sec:significance}

We can calculate the approximate difference of moving an edge
from one community to another. Assume we move an edge from community $r$ to
community $s$. The change in the density will be approximately $p_r -
\frac{1}{n_r^2}$ and $p_s + \frac{1}{n_s^2}$ respectively. The corresponding
difference in significance will be approximately
\begin{align}
  \Sig(\V') - \Sig(\V) =& 
  n_s^2 \left( D(p_s + \frac{1}{n_s^2} \midd p) - D(p_s \midd p) \right)
  \nonumber \\
  & + n_r^2 \left( D(p_r - \frac{1}{n_r^2} \midd p) - D(p_r \midd p) \right) \\
  \approx& \frac{\partial}{\partial p_s} D(p_s \midd p)
  -\frac{\partial}{\partial p_r} D(p_r \midd p) \\
  =& \log \frac{p_s}{1 - p_s} \frac{1 - p_r}{p_r} = \Delta \mathcal{Z}.
\end{align}
This quantity is particularly straightforward (the logarithmic odds ratio), and
if $p_r > p_s$ the difference will be negative, and if $p_r < p_s$ this quantity
will be positive. Moving edges from a denser community to a less dense community
decreases the significance. In other words, making two densities more equal
decreases the significance. Repeating these steps, we should expect to find the
lowest significance when the communities are of equal density.

Alternatively, by convexity of the Kullback-Leibler divergence, we obtain for
significance that
\begin{equation}
  \Sig(\V)  \geq \left(\sum_c \binom{n_c}{2}\right)D\left(\sum_c
  \frac{\binom{n_c}{2}}{\sum_c \binom{n_c}{2}} p_c \midd p\right).
\end{equation}
Realizing that $m_c = p_c \binom{n_c}{2}$, we see that 
\begin{equation}
  \sum_c \frac{\binom{n_c}{2}}{\sum_c \binom{n_c}{2}} p_c  
  = \frac{m_\text{int}}{M_\text{int}} = p\frac{q}{\langle q \rangle}.
\end{equation}
Notice that this can be interpreted as an average internal density $\langle p_c
\rangle$ as stated in the main text. Using this
we arrive at
\begin{equation}
  \Sig(\V) \geq  M_\text{int} D\left(p\frac{q}{\langle q \rangle} \midd p\right).
  \label{equ:ineq_sig}
\end{equation}
Hence, the significance of a partition with different community densities $p_c$
is generally larger than a partition where all communities have the same average
density $p_c = \frac{m_\text{int}}{M_\text{int}}$. Notice that
$p\frac{q}{\langle q \rangle}$ should be bounded by
$1$ so that $q > \langle q \rangle > p\langle q \rangle$ in general.

This points to a bound such that $\Sig(\V) \geq \Sur(\V)$ when $\langle q
\rangle < p$ in the following way. Define $\lambda  = \frac{\langle q \rangle}{p}$ so that
$\lambda < 1$ if $\langle q \rangle < p$. Again applying convexity, we obtain
\begin{align}
  \Sig(\V) &\geq M_\text{int} D(\frac{pq}{\langle q \rangle} \midd p) \\
  &= \frac{M_\text{int}}{\lambda}
    \left( \lambda D(\frac{pq}{\langle q \rangle} \midd p) + (1 - \lambda) D(0 \midd 0) \right) \\
    &\geq \frac{M_\text{int}}{\lambda} D(\lambda \frac{pq}{\langle q \rangle} \midd \lambda p) \\
    &= m_\text{int} D(q \midd \langle q \rangle) = \Sur(\V).
\end{align}

If there are fewer communities (i.e. if $p > \langle q \rangle$) the relationship
is not entirely clear, but there are cases for which surprise may be larger than
significance.  For example, if we assume an equi-sized equi-dense partition with
$r$ communities, then $q=\frac{p_c \langle q \rangle}{p}$ and $\langle q \rangle
= \frac{1}{r}$, and the difference can be written as
\begin{multline}
  \Sur(\V) - \Sig(\V) = m(1 - q)\log\frac{1-q}{1-\langle q \rangle} \\
  - M_\text{int} (1 - p_c)\log\frac{1-p_c}{1-p}.
\end{multline}
Indeed if $\langle q \rangle > p$ then $\Sur(\V) > \Sig(\V)$ for equi-sized equi-dense
partitions. Keep in mind though that an equi-sized equally dense partition will
have a lower significance in general, so that this does not hold for $\langle q
\rangle > p$ in general.

\end{document}